\documentclass[prb,twocolumn,showpacs,floats,eqsecnum,amsmath,amssymb,nofootinbib]{revtex4}
\usepackage[dvips]{graphicx}

\begin{document}



\title{Deconfinement phase transition in a two-dimensional model of
interacting $2\times 2$ plaquettes}

\author{A. Fledderjohann$^{1}$, A. Kl\"umper$^{1}$ and K.-H. M\"utter$^{1}$}

\affiliation{$^1$Physics Department, University of Wuppertal, 42097 Wuppertal,
Germany}


\begin{abstract}
\leftskip 2cm
\rightskip 2cm
A two-dimensional model of interacting plaquettes is studied by means of the
real space renormalization group approach.
Interactions between the plaquettes are mediated solely by spin excitations
on the plaquettes. Depending on the plaquette-plaquette coupling $J$, we
find two regimes:

``confinement'' $J_c< J\leq 1$, where the singlet ground state
forms an infinite (``confined'') cluster in the thermodynamical limit. Here
the singlet-triplet gap vanishes, which is the signature for long range
spin-spin correlators.

``deconfinement'' $0\leq J< J_c$, where the singlet ground state
``deconfines'' - i.e. factorizes - into finite $n$-clusters of size
$2^n\times 2^n$, with $n\leq n_c(J)$. Here the singlet-triplet gap is finite.

The critical value turns out to be $J_c=0.4822..$.
\end{abstract}

\pacs{71.10.Fd,71.27.+a,75.10.-b, 75.10.Jm}

\maketitle

\section{Introduction\label{sec1}}

\begin{figure}[ht!]
\centerline{\hspace{0.0cm}\includegraphics[width=3.5cm,angle=-90]{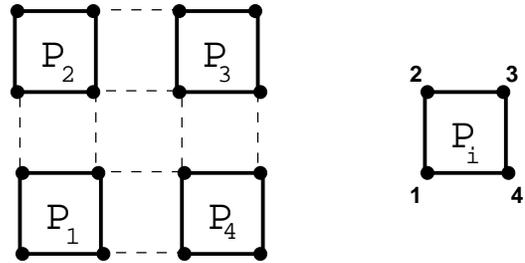}
\hspace{0.0cm}}
\caption{4-plaquette system with $2^n\times 2^n=4\times 4$ sites -- here $n=2$;
the single plaquette on the right shows the enumeration of plaquette sites.}
\label{fig_1}
\end{figure}

We will discuss in this paper the $2D$ Hamiltonian
\begin{eqnarray}
H & = & H_0+J\cdot H_J\label{h0}
\end{eqnarray}
where $H_0$ is given by isolated plaquettes occupied with spin-$1/2$
states and $H_J$ describes nearest neighbour interactions of these
plaquettes as shown in Fig. \ref{fig_1}. For $J=1$, the model 
(\ref{h0}) reduces to the well known 2$d$ antiferromagnetic Heisenberg
model. The modified Hamiltonian (\ref{h0}) has been proposed 
in studies of structural instabilities of two-dimensional systems.

The singlet-triplet gap (``spin gap'') has been studied by 
various methods:
\begin{itemize}
\item
Nonlinear $\sigma$ model as a low energy effective theory
(\onlinecite{fradkin94}), (\onlinecite{senechal93})

\item
modified spin wave theory (\onlinecite{takahashi89}), (\onlinecite{hirsch89})

\item
cluster expansion up to fourth order starting from $J=0$, i.e.
isolated plaquettes (\onlinecite{singh88})

\end{itemize}

In the spin-$1/2$ case, the model (\ref{h0}) is expected (\onlinecite{singh99}) to have a
quantum phase transition at a critical value $J_c$\footnote{
The critical value $J_c$ is related to the parameter $\gamma_c$
introduced in (\onlinecite{koga99}) by $J_c=(1-\gamma_c)/(1+\gamma_c)$}, 
which is signalled
by a vanishing singlet-triplet gap for $J>J_c$.



The results of these works are:

\begin{itemize}

\item
no phase transition for the spin $1/2$ case in the nonlinear $\sigma$ model
(\onlinecite{koga99}) for any $J\leq 1$.

\item
$J_c=0.118$ for the modified spin wave theory [versus $J_c=0.112$ in
linear spin wave theory (\onlinecite{kluemper02})].

\item
$J_c=0.54$ for the cluster expansion.

\item
A value $J_c=0.555$ was obtained by means of Ising series expansions
(\onlinecite{singh99}).

\item
$J_c$ has been determined by means of the CORE method (contractor
renormalization expansion) first by Capponi et al. (\onlinecite{capponi04})
and recently by Albuquerque et al. (\onlinecite{troyer08}). These results
are somewhat lower $J_c=0.548$.

\item
Recent Monte Carlo simulations 
(\onlinecite{aits06},\onlinecite{janke08})
yield  values close to $J_c=0.549$.

\end{itemize}

The nonlinear $\sigma$ model approach yields different results for
different cut-off schemes. In [Kawakami et al. (\onlinecite{koga99})] no phase transition
was found for $S=1/2$, whereas in [Takano et al. (\onlinecite{takano03})] a critical 
$J_c$ ($0.2\leq J_c\leq 0.25$) was obtained.

The phase transition of the magnetic system has been discussed in its
correspondence to a 
superfluid-insulator transition of the boson model (\onlinecite{senthil04}).

The authors of ref. (\onlinecite{troyer08}) start from singlet ($S=0$) 
and triplet ($S=1$) plaquette states. Excited states $|S,m\rangle$, $S=0,1,2$,
$m=-S,..,S$ are absolutely necessary to generate interactions, since singlets
alone cannot interact due to total spin conservation. 

\begin{figure}[ht!]
\centerline{\hspace{0.0cm}\includegraphics[width=7.5cm,angle=0]{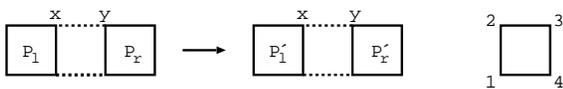}
\hspace{0.0cm}}
\caption{Interaction of neighbouring plaquettes.
}
\label{fig_2}
\end{figure}

In a recent paper
(\onlinecite{af08}), we have studied how interactions on the 4 plaquette
compound - depicted in Fig. \ref{fig_1} - are created by single plaquette
excitations.
The conservation of total spin at each interaction point is implemented
by means of the Wigner-Eckart Theorem for the transition matrix elements
\begin{eqnarray}
 & & \langle S'_l,m_l'|S_q(x)|S_l,m_l\rangle=\nonumber\\
 & & v_q\left(\begin{array}{c|cc}S'_l & 1 & S_l\\m'_l & q & m_l
\end{array}\right)M(S'_l,x,S_l)\,.\label{WET}
\end{eqnarray}
They can be expressed in terms of a Clebsch-Gordan coefficient and one 
reduced matrix element $M(S'_l,x,S_l)$. The latter only depends on the
initial and final plaquette spin $S_l,\,S'_l$ and the triplet operator
$S_q(x)$ at site $x$. The phase $v_q$ ($v_+=-1$, $v_0=v_-=1$) results
from the transformation properties of the spin operator $S_q(x)$ under
the group $SU(2)$. The interaction between neighbouring plaquettes can
be traced back to the product of reduced matrix elements at sites $x$
and $y$ (Fig. \ref{fig_2})
\begin{eqnarray}
\overline M(S'_l,S_l;S'_r,S_r) & = & \sum_{\langle x,y\rangle}
M(S'_l,x,S_l)M(S'_r,y,S_r)\nonumber\\
\end{eqnarray}

Quintuplet excitations ($S=2$) have not been considered in ref.
(\onlinecite{troyer08}). We want to stress here that triplet-quintuplet 
transitions are large - comparable with singlet-triplet transitions.
It is shown in (\onlinecite{af08}) that the gaps (singlet-triplet
and triplet-quintuplet) decrease in the renormalization process.
We will see in this paper that the inclusion of quintuplet excitations
will move the critical value $J_c$ substantially to a lower value.

The paper is organized as follows:

In Section \ref{sec2} we summarize the details of the real space
renormalization group approach in $2D$ models.

In Section \ref{sec3} we evaluate the renormalization group flow
for various couplings and gaps.

In Section \ref{sec4} we discuss the deconfinement of the ground
state wavefunction for $J<J_c$.

In Section \ref{sec5} we present a finite size analysis of the
singlet-triplet gap in both regimes: $J_c<J<1$ (confined),
$0\leq J<J_c$ (deconfined).

Section \ref{sec6} is devoted to the staggered magnetization.

\section{Real space renormalization group in $2D$ models.\label{sec2}}

In (\onlinecite{af08}) we first studied the interaction matrices $\Delta^{(2)}_S$ of the
four plaquette system (Fig. \ref{fig_1}) in the sectors with total spin
$S$. The elements of the interaction matrices are fixed on one hand by
the Clebsch-Gordan coefficients, which arise in the construction of
eigenstates with total spin $S$ (on the 4-plaquette system) and the
evaluation of the Wigner-Eckart Theorem (\ref{WET}) for the transition
matrix elements. On the other hand $\Delta^{(2)}_S$ only depends on
the following couplings
\begin{eqnarray}
\gamma & = & \frac{1}{a}\cdot\overline M(21;10)
\label{coupl2}\\
\beta & = & \frac{1}{a}\cdot\overline M(11;11)
\label{coupl3}\\
\varepsilon & = & \frac{1}{a}\cdot\overline M(22;11)
\label{coupl4}
\end{eqnarray}
and gaps
\begin{eqnarray}
\rho & = & \frac{E_1-E_0}{a}\label{scgap1}\\
\kappa & = & \frac{E_2+E_0-2E_1}{a}
\label{scgap2}
\end{eqnarray}
We have factored out from the couplings (\ref{coupl2})-(\ref{coupl4}) the
``fundamental'' interaction
\begin{eqnarray}
a & = & \overline M(1,0;1,0)\label{int_a}
\end{eqnarray}
which is induced by the singlet-triplet transitions on the plaquette.

In Appendix A of ref.\,(\onlinecite{af08})
one can find the explicit form of the interaction
matrices $\Delta_S$ (for $J=1$!) $S=0,1,2$ under the premise that on the
four plaquettes only rotational symmetric configurations with singlets,
triplets and at most one quintuplet contribute. In this case the dimensions
$d_S$ of the interaction matrices $\Delta_S$ turn out to be
\begin{eqnarray}
(d_0,d_1,d_2) & = & (7,9,14)\,.
\end{eqnarray}
The factor $J$ in (\ref{h0}) is taken into account in the interaction
matrices $\Delta_S(\frac{\rho}{J},\frac{\kappa}{J},\gamma,\beta,\varepsilon)$
by a rescaling of the normalized gaps (\ref{scgap1}), (\ref{scgap2})
whereas the couplings (\ref{coupl2})-(\ref{coupl4}) remain unchanged.

Having constructed in this way the interaction matrices $\Delta_S^{(n)}$
on an $n=2$ cluster ($2^n\times 2^n$) from the ground states on an $n=1$
cluster, we turned to the question, whether it is possible in general to
construct the interaction matrix $\Delta_S^{(n+1)}$, $S=0,1,2$ from the
corresponding quantities  of a $n\times n$ cluster. This is indeed possible
under the assumption that the low energy states on the ($n+1$)-cluster
can be built up again solely from singlet, triplet, quintuplet ground
states on $n$-clusters. The $n$-dependence only appears in a renormalization
of the couplings (\ref{coupl2})-(\ref{coupl4}) and energy differences
(\ref{scgap2}), (\ref{scgap2})

Here, we refer to (\onlinecite{af08}) $[$eqns. (6.1)-(6.4);\quad (6.5),(6.6)$]$
for the used formulas of the renormalization of the couplings and recursion
formulas for the scaled energy differences.
Note, that $J$ does not appear in the first group of equations, whereas
the remaining two (scaled energy differences) are linear in $J$.

Each step $n\rightarrow n+1$ in the renormalization procedure demands
the diagonalization of the interaction matrices $\Delta_S^{(n+1)}$,
$S=0,1,2$:
\begin{eqnarray}
\Delta_0^{(n+1)}|\sigma^{(n+1)}\rangle & = & \sigma^{(n+1)}
|\sigma^{(n+1)}\rangle\label{D0}\\
\Delta_1^{(n+1)}|\tau^{(n+1)}\rangle & = & \tau^{(n+1)}
|\tau^{(n+1)}\rangle\label{D1}\\
\Delta_2^{(n+1)}|\xi^{(n+1)}\rangle & = & \xi^{(n+1)}
|\xi^{(n+1)}\rangle\label{D2}
\end{eqnarray}
The eigenstates $|\sigma^{(n+1)}\rangle$, $|\tau^{(n+1)}\rangle$,
$|\xi^{(n+1)}\rangle$ with the largest eigenvalues $\sigma^{(n+1)}$, $\tau^{(n+1)}$,
$\xi^{(n+1)}$ enter in the quantities
\begin{eqnarray}
I^{(n+1)}(a,b) & , & G^{(n+1)}(a,b)\quad (a,b)=(1,0),(2,1)\nonumber\\
F_{\tau}^{(n+1)}(a,a) & , & F_{\xi}^{(n+1)}(a,a)\quad (a,a)=(1,1),(2,2)\nonumber
\end{eqnarray}
according to the bilinear forms:
\begin{eqnarray}
I^{(n+1)}(a,b) & = & \sum_{k,i}\tau_k^{(n+1)}I_{k,i}(a,b)\sigma_i^{(n+1)}\label{coeff}\\
G^{(n+1)}(a,b) & = & \sum_{l,k}\xi_l^{(n+1)}G_{l,k}(a,b)\tau_k^{(n+1)}\\
F_{\tau}^{(n+1)}(a,a) & = & \sum_{k}\Big(\tau_k^{(n+1)}\Big)^2F_{\tau,k}(a,a)\\
F_{\xi}^{(n+1)}(a,a) & = & \sum_{l}\Big(\xi_l^{(n+1)}\Big)^2F_{\xi,l}(a,a)\,.
\label{coeffb}
\end{eqnarray}
The contraction $I_{j,i}(1,0)$, etc. are independent of $n$ and listed
in Appendix B of paper 
(\onlinecite{af08}).

\section{Numerical evaluation of the renormalization group flow.\label{sec3}}

We now turn to the numerical evaluation of the recursion formula of the
couplings [eqns. $(6.1)-(6.4)$ in (\onlinecite{af08})]
and gaps [eqns. $(6.5),(6.6)$ in (\onlinecite{af08})] in order to study the $n$-dependence
(i.e. finite size $2^{n+1}\times 2^{n+1}$) and $J$-dependence. We start
with the singlet-triplet gap $\rho^{(n+1)}$, which yields the signature for
long range order:
From Fig. \ref{fig_3} we see, that there are two different regimes:

\begin{figure}[ht!]
\centerline{\hspace{0.0cm}\includegraphics[width=5.7cm,angle=-90]{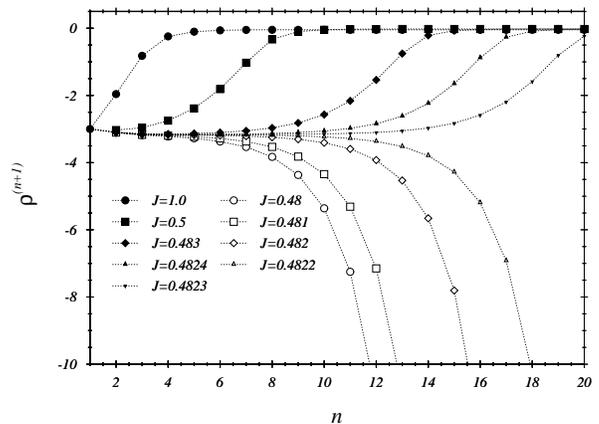}
\hspace{0.0cm}}
\caption{The scaled singlet-triplet gaps $\rho^{(n+1)}$ 
[(6.5) in (\onlinecite{af08})] as
function of $n$ and $J$.}
\label{fig_3}
\end{figure}

\begin{itemize}
\item[a)]
$J_c\leq J\leq 1$

Here the singlet-triplet gap approaches zero with increasing system size.
For $J=1$ we are close to zero already on small systems for $n_0=3$. For
decreasing $J$, $n_0(J)$ increases and seems to diverge for
$J\rightarrow J_c$.

\item[b)]
Below this critical value ($J<J_c$) the singlet-triplet gap $\rho^{(n+1)}$
does not converge to zero anymore. Note also that there is a change in the
curvature of $\rho^{(n+1)}$ with $n$, which is for large $n$ convex if
$J_c< J$ but concave if $J_c> J$. This allows for a very precise determination
of
\begin{eqnarray}
J_c & = & 0.4822..\label{crit_value}
\end{eqnarray}

\end{itemize}

\begin{figure}[ht!]
\centerline{\hspace{0.0cm}\includegraphics[width=5.7cm,angle=-90]{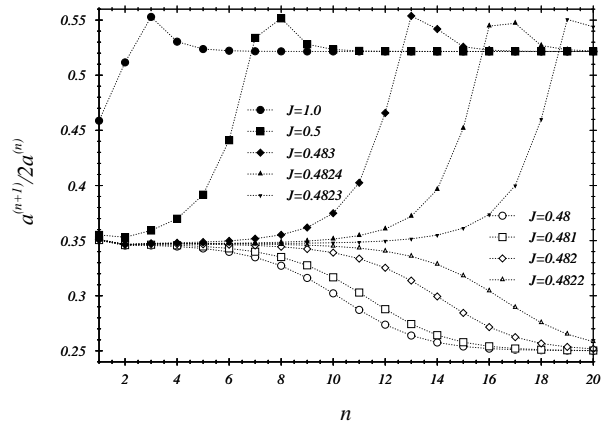}
\hspace{0.0cm}}
\caption{The coupling ratio $\frac{a^{(n+1)}}{2a^{(n)}}$ 
 [(6.1) in (\onlinecite{af08})] as
function of $n$ and $J$.}
\label{fig_4}
\end{figure}

Let us next turn to the coupling ratio  $\displaystyle
\frac{a^{(n+1)}}{2a^{(n)}}$ [(6.1) in (\onlinecite{af08})]. 
As function of $n$ this quantity has a
maximum, which travels to larger values of $n$, if $J$ is lowered
(Fig. \ref{fig_4}).

For $J_c\leq J\leq 1$ all curves approach a common limit for large $n$:
\begin{eqnarray}
\frac{a^{(n+1)}}{2a^{(n)}} & = & 0.52\,.\label{lv1}
\end{eqnarray}
For $J\leq J_c$ we observe a monotonic decrease to a limiting value,
different from (\ref{lv1}):
\begin{eqnarray}
\frac{a^{(n+1)}}{2a^{(n)}} & = & 0.25\,.\label{lv2}
\end{eqnarray}
\begin{figure}[ht!]
\centerline{\hspace{0.0cm}\includegraphics[width=5.7cm,angle=-90]{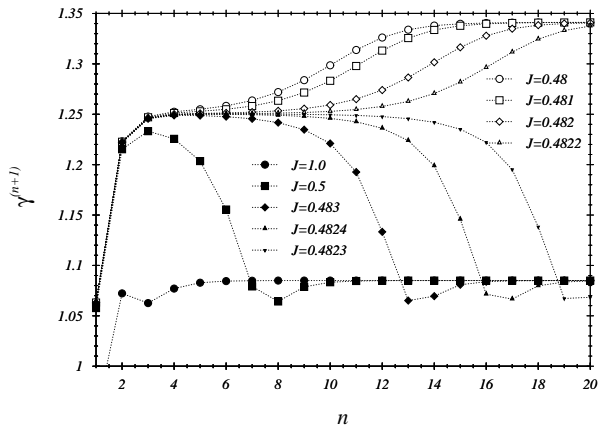}
\hspace{0.0cm}}
\caption{The coupling $\gamma^{(n+1)}$ [(6.2) in (\onlinecite{af08})] as function  of
$n$ and $J$.}
\label{fig_5}
\end{figure}

The $n$-dependence of the coupling $\gamma^{(n+1)}$ [(6.2) in (\onlinecite{af08})]
is shown in Fig. \ref{fig_5}.

For $J_c\leq J\leq 1$ all curves approach a common limit
\begin{eqnarray}
\gamma^{(n+1)} & \rightarrow & 1.0849..\label{g1}
\end{eqnarray}
for large $n$, whereas we observe a monotonic increase with $n$ for
$J<J_c$ and a common limit
\begin{eqnarray}
\gamma^{(n+1)} & \rightarrow & 1.3411..\label{g2}
\end{eqnarray}

We only want to mention that the ``diagonal'' couplings [(6.3) and (6.4) in (\onlinecite{af08})]
, which do not change the plaquette spins, die out after a
few steps.

\section{Deconfinement of the ground state wavefunction.\label{sec4}}

It was pointed out in Section \ref{sec2}, that the renormalization
group procedure demands in each step $n\rightarrow n+1$ the diagonalization
(\ref{D0})-(\ref{D2}) of the interaction matrices. We only keep those eigenvectors
($|\sigma^{(n+1)}\rangle$) with largest eigenvalue ($\sigma^{(n+1)}$).

We want to discuss now the physical meaning of the eigenvector components: 
\begin{eqnarray}
\sigma_i^{(n+1)} & =  & \langle i,0;n+1|\sigma^{(n+1)}\rangle
\quad i=1,..,7\label{sigma_i}
\end{eqnarray}
in the orthonormal basis $|i,0;n+1\rangle$ in the singlet sector -
defined in Table II of ref. (\onlinecite{af08}).
E.g. $\Big(\sigma_1^{(n+1)}\Big)^2$
has to be interpreted as the probability to find in the singlet
ground state the four plaquette configuration
\begin{eqnarray}
|1,0\rangle & = & \left(\begin{array}{cc}0 & 0\\0 & 0\end{array}\right)
\label{0000}
\end{eqnarray}
with four \underline{noninteracting}
singlets. 

\begin{figure}[ht!]
\centerline{\hspace{0.0cm}\includegraphics[width=5.7cm,angle=-90]{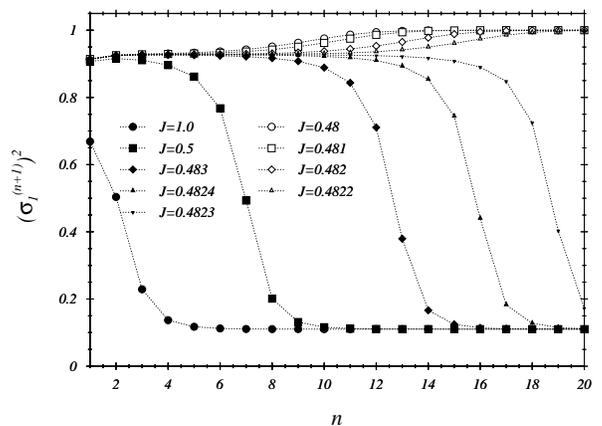}
\hspace{0.0cm}}
\caption{The probability $\Big(\sigma_1^{(n+1)}\Big)^2$ to find the
configuration (\ref{0000}) with four  noninteracting singlets in the ground state.}
\label{fig_6}
\end{figure}

If
\begin{eqnarray}
\Big(\sigma_1^{(n+1)}\Big)^2 & \rightarrow & 1\,,\label{deconf_lim_1}
\end{eqnarray}
the singlet ground state $|\sigma^{(n+1)}\rangle$ ``deconfines'' to the
configuration $|1,0\rangle$ with four noninteracting singlets.

In Fig. \ref{fig_6}, we show $\Big(\sigma_1^{(n+1)}\Big)^2$ as
function of $n$ - i.e. the system size ($2^{n+1}\times 2^{n+1}$). 

\begin{figure}[ht!]
\centerline{\hspace{0.0cm}\includegraphics[width=5.7cm,angle=-90]{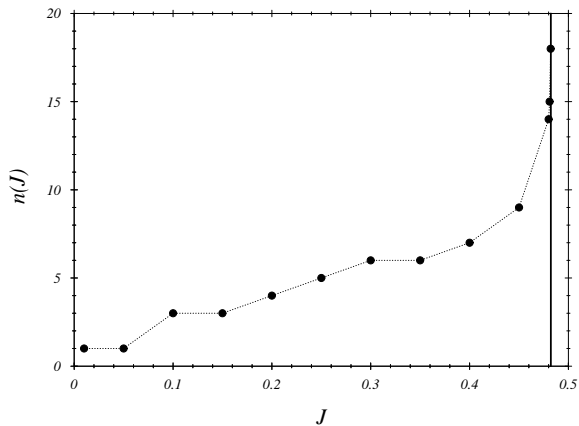}}
\caption{$n_c(J)$: as a criterion we show for each chosen $J$-value that
number of iterations ($n$) where $\Big(\sigma_1^{(n+1)}\Big)^2$
for the first time exceeds 0.999.}
\label{fig_7}
\end{figure}

For $J_c\leq J\leq 1$, $\Big(\sigma_1^{(n+1)}\Big)^2$ converges
to
\begin{eqnarray}
\Big(\sigma_1^{(n+1)}\Big)^2 & \rightarrow & 0.110236..\,.
\label{lvsig1}
\end{eqnarray}
For $J<J_c$ the deconfinement limit (\ref{deconf_lim_1}) 
is practically reached
at a finite value $n=n_c(J)$, cf. Fig. \ref{fig_7}.

$n_c(J)$ decreases with $J<J_c$ and defines the largest
cluster size ($2^{n_c(J)}\times 2^{n_c(J)}$) which is still confined.

In summary we can say: In the two-dimensional system of interacting
plaquettes defined in (\ref{h0}) we observed two phases.
\begin{itemize}
\item
For $J_c\leq J\leq 1$, there is a confined phase, where the
ground state does not factorize into finite $n$-clusters 
($2^n\times 2^n$) but forms one infinite cluster in the 
thermodynamical limit $n\rightarrow\infty$. The vanishing
of the singlet-triplet gap is the characteristic signature
of this phase.

\item
In the deconfined phase $0\leq J\leq J_c$ the ground state
factorizes into finite clusters $n<n_c(J)$ where $n_c(J)$
defines the maximal size of clusters and is shown in Fig. 
\ref{fig_7}.


\end{itemize}

\section{Finite-size analysis of the singlet-triplet gap.\label{sec5}}

\begin{figure}[ht!]
\centerline{\hspace{0.0cm}\includegraphics[width=5.7cm,angle=-90]{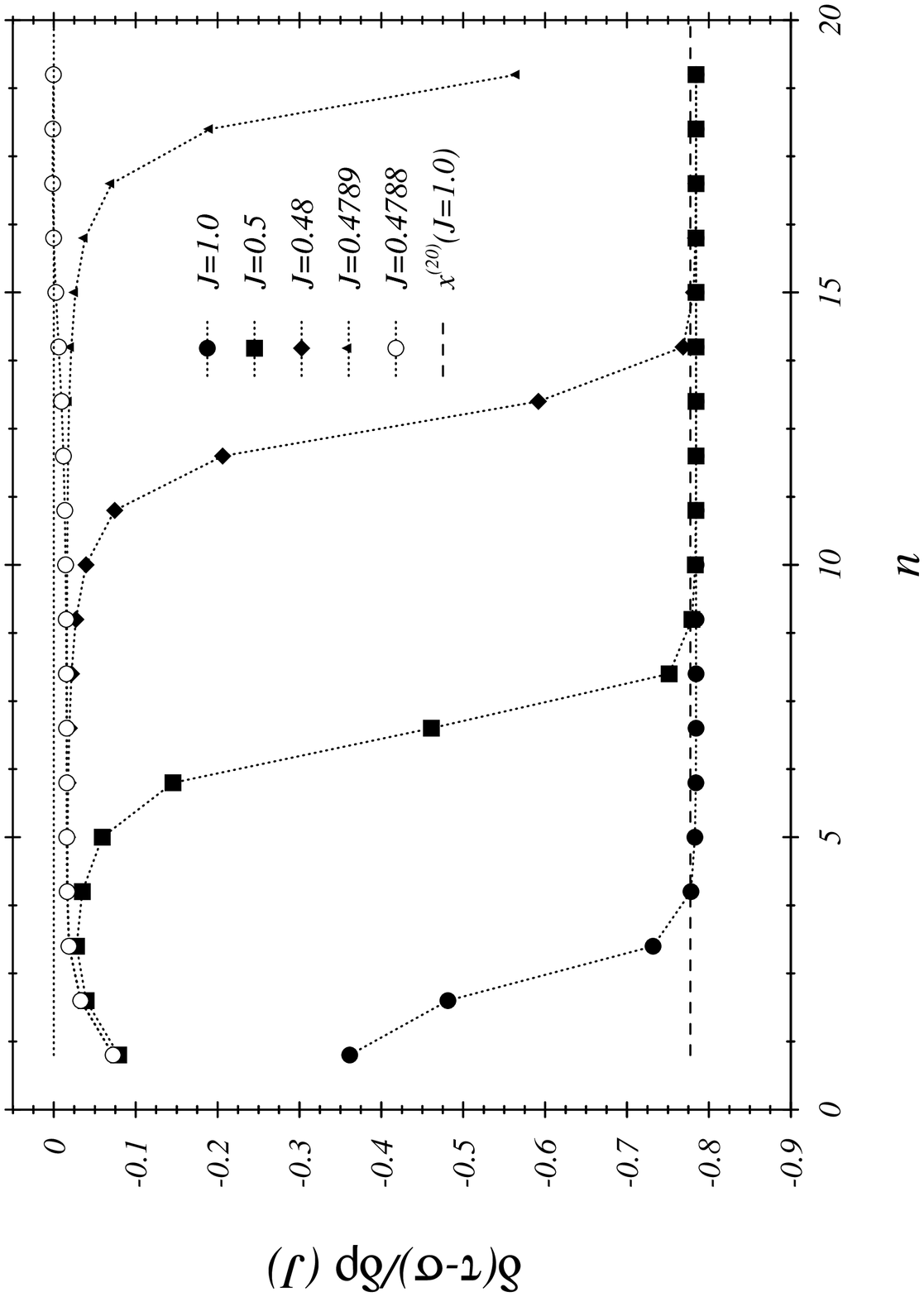}}
\caption{The first derivative [(\ref{x_value}), (\ref{f-deriv})] with respect to
$\rho$.}
\label{fig_8}
\end{figure}

In the confined regime $J_c\leq J\leq 1$ the singlet-triplet gap
\begin{eqnarray}
E_1^{(n)}-E_0^{(n)} & \sim & 4^{-n\nu_1}
\end{eqnarray}
vanishes with an exponent
\begin{eqnarray}
\nu_1 & = & -\frac{\log(1+x)}{\log 4}\label{universal_exp}
\end{eqnarray}
which can be determined from the first derivative
\begin{eqnarray}
x = \frac{d(\tau-\sigma)}{d\rho} & = & 
\frac{\partial(\tau-\sigma)}{\partial\rho}+
\frac{\partial(\tau-\sigma)}{\partial\kappa}\cdot\frac{d\kappa}{d\rho}
\label{x_value}\hspace{0.6cm}
\end{eqnarray}
of the largest eigenvalues $\sigma$, $\tau$ of the interaction matrices
$\Delta_S(\rho,\kappa,...)$ in the singlet ($S=0$) and triplet ($S=1$)
sector:
\begin{eqnarray}
\frac{\partial(\tau-\sigma)}{\partial\rho} & = & \langle\tau|\frac{\partial\Delta_1}
{\partial\rho}|\tau\rangle-\langle\sigma|\frac{\partial\Delta_0}
{\partial\rho}|\sigma\rangle\nonumber\\
 & = & -4(\tau_1^2-\sigma_1^2)-2(\tau_2^2+\tau_3^2+\tau_4^2+\tau_5^2)\nonumber\\
 & & +2(\sigma_2^2+\sigma_3^2)\label{f-deriv}\\
\frac{\partial(\tau-\sigma)}{\partial\kappa} & = & \langle\tau|\frac{\partial\Delta_1}
{\partial\kappa}|\tau\rangle-\langle\sigma|\frac{\partial\Delta_0}
{\partial\kappa}|\sigma\rangle\nonumber\\
 & = & 1-\tau_1^2-\tau_2^2-\tau_3^2-\sigma_6^2-\sigma_7^2\label{f-deriv2}\,.
\end{eqnarray}
The $n$- and $J$-dependence of $\frac{\partial(\tau-\sigma)}{\partial\rho}$ 
(\ref{f-deriv}) - which is the dominant part of (\ref{x_value}) -
 is presented
in Fig. \ref{fig_8}. 
In the confined regime $J_c\leq J\leq 1$ all the
curves approach a common limit
\begin{eqnarray}
x^{(n+1)}(J) & \rightarrow & -\frac{3}{4}
\end{eqnarray}
which leads to a universal exponent (\ref{universal_exp})
\begin{eqnarray}
\nu_1(J) & = & 1\,.
\end{eqnarray}

\begin{figure}[ht!]
\centerline{\hspace{0.0cm}\includegraphics[width=5.7cm,angle=-90]{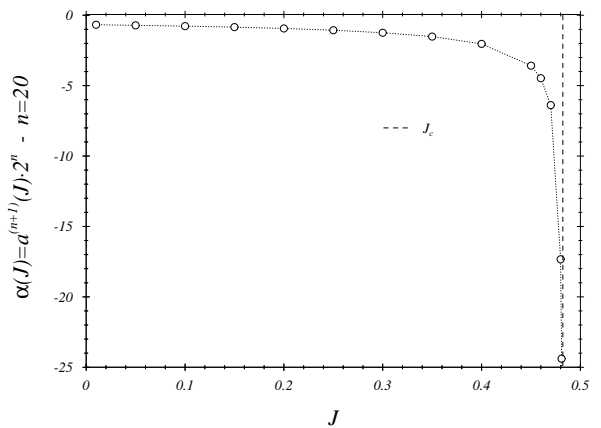}}
\caption{$n$-dependence of $\alpha(J)=2^{-n}\cdot a^{(n+1)}(J)$
- shown after $n=20$ iterations.}
\label{fig_9}
\end{figure}

In the deconfined regime $0\leq J<J_c$ we find a nonvanishing
singlet-triplet gap:
\begin{eqnarray}
E_1^{(n)}-E_0^{(n)} & = & E_1^{(\infty)}-E_0^{(\infty)}+f^{(n)}
\label{f-s_cor}
\end{eqnarray}
with a finite-size correction
\begin{eqnarray}
f^{(n+1)}-f^{(n)} & = & J\cdot a^{(n)}\Big(\tau^{(n+1)}-\sigma^{(n+1)}\Big)
\end{eqnarray}
which follows from the difference $\tau^{(n+1)}-\sigma^{(n+1)}$ of
the largest eigenvalues $\tau^{(n+1)}$, $\sigma^{(n+1)}$ in the
triplet and singlet sector and the fundamental coupling $a^{(n)}$
(\ref{int_a}), which can be extracted from Fig. \ref{fig_4}.
The large $n$ limit of the difference
\begin{eqnarray}
\tau^{(n+1)}-\sigma^{(n+1)} & \rightarrow & 2
\end{eqnarray}
turns out to be 2 for all $J$ $0<J<J_c$ whereas the coupling
\begin{eqnarray}
a^{(n+1)}(J) & = & \alpha(J)\cdot 2^{-n}
\end{eqnarray}
decreases with the system size $N=4^n$ as $\frac{1}{\sqrt{N}}$ for
all $J$ $0\leq J<J_c$. $\alpha(J)$ is shown in Fig. \ref{fig_9}.

\section{The staggered magnetization.\label{sec6}}

Finally we want to present our results from the recursion formula
[(8.3) in ref. (\onlinecite{af08})]

\begin{eqnarray}
R^{(n+1)} & = & \frac{\langle\sigma^{(n+1)}|\Sigma_-^{(n+1)}\Sigma_+^{(n+1)}|
\sigma^{(n+1)}\rangle}{\langle\sigma^{(n)}|\Sigma_-^{(n)}\Sigma_+^{(n)}|
\sigma^{(n)}\rangle}\nonumber\\
 & = & \sum_{i',i=1}^7\sigma_{i'}^{(n+1)}\sigma_i^{(n+1)}\Gamma_{i',i}
(\gamma^{(n)})
\end{eqnarray}
for the staggered magnetization on an ($n+1$)-cluster. Note, that the
renormalization procedure only enters via the components $\sigma_i^{(n+1)}$,
$i=1,\ldots,7$ on an ($n+1$)-cluster and the coupling $\gamma^{(n)}$.

The $7\times 7$ matrix $\Gamma_{i',i}(\gamma^{(n)})$ is presented in
Appendix C of (\onlinecite{af08}). In Fig. \ref{fig_10} we present
the ratio $R^{(n+1)}$ as function of $n$ and $J$ for the case ($d_0=7,
d_1=9,d_2=14$) for large $n$; 

\begin{figure}[ht!]
\centerline{\hspace{0.0cm}\includegraphics[width=5.7cm,angle=-90]{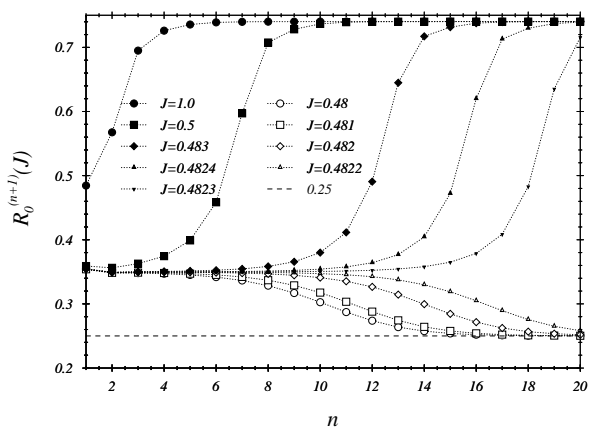}
\hspace{0.0cm}}
\caption{The ratio $R^{(n+1)}$ as function of $n$ and $J$ for
dimensions: $d_0=7$, $d_1=9$, $d_2=14$.}
\label{fig_10}
\end{figure}
%
all the curves approach a common limit
\begin{eqnarray}
R^{(n)}\rightarrow 0.7401.. & & \mbox{for }J_c\leq J\leq 1\\
R^{(n)}\rightarrow 0.25\hspace{0.5cm} & & \mbox{for }0\leq J\leq J_c\nonumber
\end{eqnarray}

\section{Discussion and perspectives.\label{sec7}}

We have studied in the $2D$ model with interacting plaquettes various 
observables like the scaled singlet-triplet gap $\rho^{(n+1)}$
(Fig. \ref{fig_3}) as function of $n$ (i.e. system size $2^{n+1}\times 2^{n+1}$)
and the coupling parameter $J$ in (\ref{h0}). We find spectacular differences
in the confinement ($J_c<J\leq 1$) and deconfinement ($J<J_c$) regime, which
allows - for a given truncation scheme -
for an extremely precise determination of the critical coupling $J_c$
in all these quantities. This means, that the interaction matrices
$\Delta_S^{(n+1)}$, $S=0,1,2$ (\ref{D0})-(\ref{D2}) and thereby the
renormalization group equations [(6.2)-(6.6) in (\onlinecite{af08})] depend on $J$ in
an extremely sensitive way. The reason is a feedback between the scaled
energy differences $\rho^{(n)},\,\kappa^{(n)}$ - which enter in the diagonals
of $\Delta_S^{(n+1)}$ - and the largest eigenvalues $\sigma^{(n+1)}$, $\tau^{(n+1)}$,
$\xi^{(n+1)}$ (\ref{D0})-(\ref{D2}).

This feedback also leads to a dramatic change in the eigenstates
$|\sigma^{(n+1)}\rangle$, $|\tau^{(n+1)}\rangle$, $|\xi^{(n+1)}\rangle$.
E.g. the square of the first component $[$(\ref{sigma_i}) for
$i=1]$ $\Big(\sigma_1^{(n+1)}\Big)^2$ in the singlet eigenvector
$|\sigma^{(n+1)}\rangle$ changes completely if we go from the
confined ($J_c\leq J\leq 1$) to the deconfined ($J<J_c$) regime.
In the deconfined regime $\Big(\sigma_1^{(n+1)}\Big)^2$ is almost one, which
means, that the ground state factorizes into 4 noninteracting
singlets. In the confined phase $\Big(\sigma_1^{(n+1)}\Big)^2$ is very small.
Therefore, the remaining components $\sigma_i^{(n+1)}$, $i=2,..,7$
contribute significantly to the eigenstate $|\sigma^{(n+1)}\rangle$.

These contributions are characterized by excitations of the cluster spins
on the four plaquette system. Excitations of cluster
spins are necessary to induce cluster-cluster interactions. 
The vanishing of the singlet-triplet gap - as it is observed in the
confinement regime $J_c\leq J\leq 1$ - is a consequence of the
cluster-cluster interactions induced by cluster excitations (triplet
and quintuplet). We have checked the dependence on the truncation of
the interaction matrix by suppressing in Tables II, III, IV (of ref.
(\onlinecite{af08})) all states with one quintuplet plaquette. The
dimensions of the interaction matrices reduce to
\begin{eqnarray}
(d_0,d_1,d_2) & = & (5,3,4)\label{sig1_plot}
\end{eqnarray}
which of course worsens the renormalization group approach. This is
signalled by a somewhat larger singlet-triplet gap. As a consequence
the deconfined regime ($J\leq J_c$) is enlarged.

If we look at the deconfinement parameter $\Big(\sigma_1^{(n+1)}\Big)^2$,
Fig. \ref{fig_11} for the case (\ref{sig1_plot}), we observe a shift to a larger
value of $J_c$:

\begin{figure}[ht!]
\centerline{\hspace{0.0cm}\includegraphics[width=5.7cm,angle=-90]{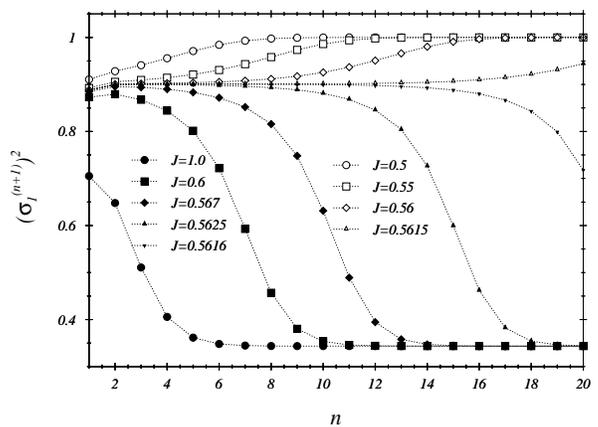}
\hspace{0.0cm}}
\caption{The probability $\Big(\sigma_1^{(n+1)}\Big)^2$ evaluated for
dimensions (\ref{sig1_plot}).}
\label{fig_11}
\end{figure}

\begin{eqnarray}
J_c(5,3,4)=0.5615.. & ; & J_c(7,9,14)=0.4822..\,.\label{Jc_neu}
\end{eqnarray}
This value is close to the result of ref. (\onlinecite{troyer08})
obtained without quintuplet excitations. Therefore, the difference
in the two values (\ref{Jc_neu}) reflects the effect of rotational
symmetric excited states on the 4 plaquette cluster with one
quintuplet. We expect that further excited states with
$n_Q=2,3,4$ quintuplets will lead to changes in the values $J_c$
as well. The Monte Carlo simulations of Janke et al. (\onlinecite{janke08})
suggest, that the RG results should converge non-monotonously
towards $0.549$.


\begin{thebibliography}{99}





\bibitem{fradkin94}
E. Fradkin, {\em Field Theory of condensed matter physics},
Addison Wesley (1994);
A. M. Tsvelik, {\em Quantum Field Theory in Condensed Matter Physics},
Cambridge University press, New York (1995)

\bibitem{senechal93}M. S\'en\'echal, Phys. Rev. B {\bf 47}, 8353 (1993);
Phys. Rev. B {\bf 48}, 15890 (1993)


\bibitem{takahashi89}M. Takahashi, Phys. Rev. B {\bf 40}, 2494 (1989)


\bibitem{hirsch89}J. E. Hirsch, S. Tang, Phys. Rev. B {\bf 40}, 4769 (1989)

\bibitem{singh88}R. P. Singh, M. P. Gelfand, A. Huse, Phys. Rev. Lett. {\bf 61},
2484 (1988); H. K. He, C. J. Hamer, J. Oitmaa, J. Phys. A {\bf 23}, 1775 (1990);
K. Hida, J. Phys. Soc. Jpn. {\bf 61}, 1013 (1992)

\bibitem{kluemper02}J. Sirker, A. Kl\"umper, K. Hamacher, 
Phys. Rev. B {\bf 65}, 134409 (2002)


\bibitem{singh99}R. P. Singh, Z. Weihong, C. J. Hamer and J. Oitma, 
Phys. Rev. B {\bf 60}, 7278 (1999)

\bibitem{koga99}A. Koga, S. Kumada, N. Kawakami, 
J. Phys. Soc. Jpn. {\bf 68}, No. 2, 1999;
J. Phys. Soc. Jpn. {\bf 68}, No. 7, 1999
%

\bibitem{capponi04}S. Capponi, A. L\"auchli and M. Mambrini,
Phys. Rev. B {\bf 70}, 1004424 (2004);
S. Capponi, Theor. Chem. Acc., {\bf 116}, 524 (2006)

\bibitem{troyer08}A. F. Albuquerque, M. Troyer, J. Oitmaa,
Phys. Rev. B\,{\bf 78}, 132402 (2008)

\bibitem{aits06}C. H. Aits, U. L\"ow, A. Kl\"umper, W. Weber,
Phys. Rev. B {\bf 74}, 014425 (2006)

\bibitem{janke08}S. Wenzel, L. Bogacz, W. Janke,
Phys. Rev. Lett. {\bf 101}, 127202 (2008);
S. Wenzel, W. Janke,
Phys. Rev. B {\bf 78}, 064402 (2008), 099902(E) (2008)

\bibitem{takano03}K. Takano, Y. Kito, Y. \={O}no, and K. Sano,
Phys. Rev. Lett. \bf{91}, 197202 (2003)

\bibitem{senthil04}T. Senthil et al.,Phys. Rev. B 70, 144407 (2004)



\bibitem{af08} A. Fledderjohann, A. Kl\"umper, K.-H. M\"utter,
subm. to Eur. Phys. J. (2008)

\end{thebibliography}
\end{document}